\documentclass[12pt]{article}

\usepackage[cp1251]{inputenc}
\usepackage[english]{babel}
\usepackage[arrow,matrix]{xy}
\usepackage{amsfonts}
\usepackage{amsmath}
\usepackage{amssymb}
\usepackage{latexsym}
\usepackage{amscd}

\newtheorem{theorem}{Theorem}
\newtheorem{lemma}[theorem]{Lemma}
\newtheorem{prop}[theorem]{Proposition}

\newcommand{\rk}{{\rm rk}}
\newcommand{\id}{{\rm id}}
\newcommand{\Aut}{{\rm Aut}}
\newcommand{\qpr}{{\rm qpr}}
\newcommand{\cO}{{\cal O}}
\newcommand{\cS}{{\cal S}}

\newcommand{\normaleq}{\trianglelefteq}

\newcommand{\fe}{\varphi}

\newcommand{\lra}{\longrightarrow}
\newcommand{\ov}{\overline}

\newcommand{\wt}{\widetilde}
\newcommand{\sse}{\subseteq}
\newcommand{\lu}{{\langle}}
\newcommand{\ru}{{\rangle}}

\begin{document}

\title{On symmetries of the Strassen algorithm}
\author{Vladimir P.Burichenko}
\date{}
\maketitle

\begin{center}
Institute of mathematics of Academy of Sciences of Belarus \\
Kirov Street 32a, Gomel 246000, Republic of Belarus \\
vpburich@gmail.com
\end{center}

\begin{abstract}
We consider the famous Strassen algorithm for fast multiplication
of matrices. We show that this algorithm has a nontrivial finite
group of automorphisms of order 36 (namely the direct product of
two copies of the symmetric group on 3 symbols), or even 72, if we
consider ``extended'' Strassen algorithm. This is an indirect
evidence that the (unknown at present) optimal algorithm for
multiplication of two size 3 by 3  matrices also may have a large
automorphisms group, and this may be a fruitful idea for a search
of such an algorithm. In the beginning we give a brief
introduction to the subject, to make the text accessible for
specialists in the representation theory of finite groups.

\end{abstract}

\paragraph{1. Strassen algorithm.} This text deals with the fast
algorithms for matrix multiplication.

The usual algorithm (``multiplying of a row by a column'') needs $N^3$
multiplications and $N^3-N^2$ additions to multiply two $N\times N$
matrices, total $O(N^3)$ arithmetical operations. In 1968 V.Strassen
\cite{Strassen} discovered another approach, which is now widely known.

Let $R$ be a (non-commutative) ring,
$$ A=\begin{pmatrix} a_{11} & a_{12} \\ a_{21} & a_{22} \end{pmatrix}, \qquad
B=\begin{pmatrix} b_{11} & b_{12} \\ b_{21} & b_{22} \end{pmatrix} $$
be two $2\times 2$ matrices over $R$, and let
$$ C=\begin{pmatrix} a_{11}b_{11}+ a_{12}b_{21} & a_{11}b_{12}+a_{12}b_{22} \\
a_{21}b_{11}+a_{22}b_{21} &  a_{21}b_{12}+a_{22}b_{22} \end{pmatrix} =
\begin{pmatrix} c_{11} & c_{12} \\ c_{21} & c_{22} \end{pmatrix} $$
be their product. Then we can compute $C$ with only 7 multiplications
in $R$ (but at price of 18 additions).
Namely, consider the following products:
$$ p_1=a_{11}(b_{12}+b_{22}), \quad p_2=(a_{11}-a_{12})b_{22},
\quad p_3=(-a_{21}+a_{22})b_{11}, $$
$$ p_4=a_{22}(b_{11}+b_{21}), \quad p_5=(a_{11}+a_{22})(b_{11}+b_{22}), $$
$$ p_6=(a_{11}+a_{21})(b_{11}-b_{12}),\quad p_7=(a_{12}+a_{22})(b_{21}-b_{22}). $$
Then it is easy to see that
$$ c_{11}=-p_2-p_4+p_5+p_7, \quad c_{12}=p_1-p_2, \quad c_{21}=-p_3+p_4, $$
$$ c_{22}=-p_1-p_3+p_5-p_6. $$

Now let $A$ and $B$ be two $2^n\times 2^n$ matrices over $R$. We
consider them as $2\times 2$ matrices over the ring
$S=M_{2^{n-1}}(R)$. So one can multiply $A$ and $B$ using $7$
multiplications and $18$ additions in $S$. Applying the same
method recursively to compute products in $S$, we can see that it
suffices $7^n$ multiplications and $6(7^n-4^n)$ additions in $R$
to multiply $A$ and $B$.

Further, let $N\geq1$ be arbitrary natural number and take the
smallest $k$ such that $2^k\geq N$. Clearly, the number of
operations needed to multiply two $N\times N$ matrices does not
exceed such a number for $2^k\times 2^k$ matrices. Therefore we
can multiply two $N\times N$ matrices using
$O(N^{\log_27})=O(N^{2.81})$ arithmetical operations.

\paragraph{2. Motivation of the work.} It is natural that Strassen's idea
has lead to numerous investigations in various directions. The
author would like to mention survey \cite{Landsberg} and lecture
notes \cite{Blaser-lectures}. The best estimate for the number of
operations known at the moment is $O(N^{2.323})$; it is mentioned
in a talk of V.Strassen \cite{Strassen-talk}. However, the
constants in such estimates are large, and the corresponding
algorithms are mainly of theoretical interest.

On the other hand, it is not much known about the complexity of multiplication
of matrices of given ``small'' formats. Let $R(m,n,p)$ be the number of
multiplications needed to multiply $m\times n$ matrix (over a {\em non-commutative}
algebra) by a $n\times p$
matrix. It is known that $R(m,n,p)$ is symmetric in $m,n,p$ and that
\\ $R(2,2,2)=7$ \cite{Winograd},
\\ $R(2,2,3)=11$ \cite{Alexeev},
\\ $R(2,2,4)=14$ \cite{Al-Sm},
\\ $14\leq R(2,3,3)\leq 15$ (see \cite{Blaser} and \cite{Hop-Kerr} for lower and upper
estimates, respectively), and
\\ $19\leq R(3,3,3)\leq 23$ (\cite{Blaser} and \cite{Laderman}, respectively).

The author believes that at the moment it is important, both from theoretical
and practical viewpoint, to find the precise value of $R(3,3,3)$, and,
moreover, to determine the variety of optimal (i.e. with minimal number
of multiplications) algorithms in the cases $(m,n,p)=(2,2,3)$, $(2,2,4)$,
$(2,3,3)$, $(3,3,3)$
(for $(2,2,2)$ this variety is determined in \cite{Groote}, see the details below).

Now we can describe the motivation for the present work as follows.
\begin{itemize}
\item We show that the Strassen algorithm has rather large (of order 36 or even
72, depending on the viewpoint) group of automorphisms;
\item this is an indirect evidence that the (unknown at present) optimal
algorithm for the case $(m,n,p)=(3,3,3)$ has a large group of automorphisms also,
\end{itemize}
and therefore
\begin{itemize}
\item to assume in advance that the above mentioned optimal algorithm
{\em must} have a large group of automorphisms may be a fruitful idea in the
{\em search} of such an algorithm; this may be a good problem for
specialists in finite linear groups.
\end{itemize}

\paragraph{3. Decomposable tensors, rank, and Segre isomorphisms.} Let
$$ \wt V=V_1\otimes\ldots\otimes V_n $$
be the tensor product of several spaces over a field~$K$. The elements
of $\wt V$ of the form $v_1\otimes \ldots\otimes v_n$, $v_i\in V_i$,
are called {\em decomposable tensors}. For an element $t\in\wt V$
let $\rk(t)=r$ be the minimal number such that $t=t_1+\ldots+t_r$,
where $t_1,\ldots, t_r$ are decomposable. Thus, the tensors of rank
1 are precisely the decomposable tensors.

The set of all decomposable tensors in $\wt V$ is a cone, closed in the
Zariski topology. We shall call it the {\em Segre variety} (usually,
however, this term is used for the {\em projectivization} of this cone).

Let $\wt U=U_1\otimes \ldots \otimes U_m$ be another product of spaces.
Then by a {\em Segre isomorphism} we mean an isomorphism of linear spaces
$\fe:\wt U\lra\wt V$ which maps the Segre variety in $\wt U$ bijectively
to the Segre variety in $\wt V$.

If $m=n$, $\tau$ an arbitrary permutation on $\{1,\ldots,n\}$ such that
$\dim U_i=\dim V_{\tau(i)}$, $i=1,\ldots,n$, and $\alpha_i:U_i\lra
V_{\tau(i)}$ are isomorphisms, then there exists a unique isomorphism
$\fe:\wt U\lra\wt V$ such that
$$ \fe(u_1\otimes\ldots\otimes u_n)=\alpha_{\tau^{-1}(1)}(u_{\tau^{-1}(1)})\otimes
\ldots \otimes \alpha_{\tau^{-1}(n)} (u_{\tau^{-1}(n)}), \qquad \forall\: u_i\in U_i\,,$$
and this isomorphism is a Segre isomorphism.

\begin{prop}    \label{prop:Segre}
If the field $K$ is infinite and $\dim U_i$, $\dim V_j >1$ for all $i$, $j$,
then any Segre isomorphism $\fe:\wt U\lra\wt V$ has the described form.
\end{prop}

{\em Proof.} First note the following.  We say that two decomposable
tensors
$$ u=u_1\otimes\ldots\otimes u_m,\qquad v=v_1\otimes\ldots\otimes v_m\in\wt U$$
are {\em adjacent}, if $\langle u\rangle\ne \langle v\rangle$ and
$\langle u_i\rangle\ne \langle v_i\rangle$ for exactly one index $i$.
It is easy to see that $u$ and $v$ are adjacent if and only if
$\langle u\rangle\ne \langle v\rangle$ and each element of $\langle u,v\rangle$
is decomposable.

Now suppose that $\fe:\wt U\lra\wt V$ is a Segre isomorphism. It follows from
the previous remark that $\fe$ takes adjacent tensors to adjacent ones.

Take arbitrary element $u=u_1\otimes\ldots\otimes u_m\in\wt U$. Let
$v=\fe(u)$, $v=v_1\otimes\ldots\otimes v_n\in\wt V$.
The set of all $w$ adjacent to $u$ coincides with the union of subspaces
$$ R_i=u_1\otimes\ldots\otimes u_{i-1}\otimes U_i\otimes u_{i+1}
\otimes\ldots\otimes u_m\,.$$
If $i\ne j$, then $R_i\cap R_j=\langle u\rangle$. Similarly, the set
of all $z$, adjacent to $v$, coincides with the union of subspaces
$$ L_j=v_1\otimes\ldots\otimes v_{j-1}\otimes V_j\otimes v_{j+1}
\otimes\ldots\otimes v_n\,.$$

Since $\fe$ is a Segre isomorphism, it maps $\cup_{i=1}^m R_i$ bijectively
to $\cup_{j=1}^n L_j$. Since $R_i\cap R_j=\langle u\rangle$ and
$L_i\cap L_j=\langle v\rangle$ when $i\ne j$, it follows that $m=n$
and there exists a permutation $\tau$ such that $\fe(R_i)=L_{\tau(i)}$,
whence $\dim R_i=\dim L_{\tau(i)}$, $i=1,\ldots,m$. Therefore,
$\dim U_i=\dim V_{\tau(i)}$.

We may assume without loss of generality (the details are left to the
reader), that $\tau$ is the identity. Therefore, there exist isomorphisms
$\alpha_i:U_i\lra V_i$ such that $\fe$ maps
$$ u_1\otimes\ldots\otimes u_{i-1}\otimes u\otimes u_{i+1}
\otimes\ldots\otimes u_n$$
to
$$ v_1\otimes\ldots\otimes v_{i-1}\otimes \alpha_i(u)\otimes v_{i+1}
\otimes\ldots\otimes v_n\,,$$
for all $u\in U_i$.

It remains to prove that
$$ \fe(t_1\otimes\ldots\otimes t_n)=\alpha_1(t_1)\otimes\ldots\otimes
\alpha_n(t_n) \qquad \forall\ (t_1,\ldots,t_n)\in U_1\times\ldots\times U_n.$$
We argue by induction on the number $l$ of indices $i$ such that
$\langle t_i\rangle\ne \langle u_i\rangle$.

Notice that $\alpha_i(u_i)=v_i$ for all $i=1,\ldots,n$, so for $l=0$
the claim is true. For $l=1$ it is true also, by previous consruction.

Suppose $l=2$; for example, let $\langle t_1\rangle\ne \langle u_1\rangle$,
$\langle t_2\rangle\ne \langle u_2\rangle$, and $t_i=u_i$ for $i\geq3$.
The tensors $t_1\otimes u_2\otimes u_3\otimes\ldots\otimes u_n$ and
$u_1\otimes t_2\otimes u_3\otimes\ldots\otimes u_n$ are adjacent to
$t_1\otimes t_2\otimes u_3\otimes\ldots\otimes u_n$. So
$\fe(t_1\otimes t_2\otimes u_3\otimes\ldots\otimes u_n)$ is adjacent to
$$\fe(t_1\otimes u_2\otimes \ldots \otimes u_n)=\alpha_1(t_1)\otimes v_2
\otimes\ldots\otimes v_n$$
and to
$$\fe(u_1\otimes t_2\otimes u_3\otimes\ldots \otimes u_n)=v_1\otimes
\alpha_2(t_2)\otimes v_3\otimes\ldots\otimes v_n\,.$$
Therefore, $\fe(t_1\otimes t_2\otimes u_3\otimes \ldots
\otimes u_n)$
is proportional to
$$\alpha_1(t_1)\otimes \alpha_2(t_2)\otimes v_3\otimes \ldots
\otimes v_n\,.$$ As this is true for all $t_1\otimes t_2\in
U_1\otimes U_2$, the proportionality coefficient does not depend
on $t_1, t_2$, and so equals $1$.

The same argument works for other cases with $l=2$.

Finally, let $l\geq3$, for example let $\langle t_1\rangle\ne \langle u_1\rangle$,
$\langle t_2\rangle\ne \langle u_2\rangle$, $\langle t_3\rangle\ne \langle u_3\rangle$,
and $t_i=u_i$ for $i\geq4$. Then it suffices to observe that
$$\fe(t_1\otimes t_2\otimes t_3\otimes u_4\otimes \ldots \otimes u_n)$$
must be adjacent to
$$\fe(u_1\otimes t_2\otimes t_3\otimes u_4\otimes \ldots \otimes u_n)
=v_1\otimes \alpha_2(t_2)\otimes \alpha_3(t_3)\otimes v_4\otimes \ldots \otimes v_n$$
and to
$$\fe(t_1\otimes u_2\otimes t_3\otimes u_4\otimes \ldots \otimes u_n)
=\alpha_1(t_1)\otimes v_2\otimes \alpha_3(t_3)\otimes v_4\otimes\ldots\otimes v_n\,.$$
The details are left to the reader. $\square$
\medskip

We need another important concept.
Let $t\in\wt V$ be an arbitrary tensor. The {\em isotropy group} $\Gamma(t)$
is the group of all automorphisms $\wt A$ of $\wt V$ of the form
$\wt A=A_1\otimes \ldots\otimes A_n$, $A_i\in GL(V_i)$, such that $\wt A(t)=t$.
The {\em extended isotropy group} $\wt\Gamma(t)$ is the group of all
Segre automorphisms $\fe$ of $\wt V$
such that $\fe(t)=t$. Hence $\Gamma(t)\normaleq \wt\Gamma(t)$, and
$\wt\Gamma(t)/\Gamma(t)$ is a subgroup in $S_n$ (where $S_n$ is the
symmetric group on $n$ letters).

\paragraph{4. Algorithms and decomposable tensors.} Note that the Strassen
algorithm may be expressed by the following formula:
\begin{eqnarray}
\begin{pmatrix} x_{11} & x_{12} \\ x_{21} & x_{22} \end{pmatrix}
\begin{pmatrix} y_{11} & y_{12} \\ y_{21} & y_{22} \end{pmatrix}
&=& x_{11}(y_{12}+y_{22}) \begin{pmatrix} 0 & 1 \\ 0 & -1 \end{pmatrix} \label{f:1}   \\
&+& (x_{11}-x_{12})y_{22} \begin{pmatrix} -1 & -1 \\ 0 & 0 \end{pmatrix}
+(-x_{21}+x_{22})y_{11} \begin{pmatrix} 0 & 0 \\ -1 & -1 \end{pmatrix} \nonumber\\
&+& x_{22}(y_{11}+y_{21})\begin{pmatrix} -1 & 0 \\ 1 & 0 \end{pmatrix}
+ (x_{11}+x_{22})(y_{11}+y_{22})\begin{pmatrix} 1 & 0 \\ 0 & 1 \end{pmatrix} \nonumber \\
&+& (x_{11}+x_{21})(-y_{12}+y_{11})\begin{pmatrix} 0 & 0 \\ 0 & -1 \end{pmatrix}
+ (x_{12}+x_{22})(y_{21}-y_{22})\begin{pmatrix} 1 & 0 \\ 0 & 0 \end{pmatrix}. \nonumber
\end{eqnarray}

The formulae of this kind are related to decompositions of the
structure tensors of maps.

Below $K$ denotes an infinite field of characteristic $0$, and $R$ a (non-commutative)
algebra over $K$.

Recall that to a bilinear map $f:U\times V\lra W$, where $U$, $V$ and
$W$ are spaces, there corresponds a {\em structure tensor}
$\ov f\in U^\ast\otimes V^\ast\otimes W$. For example, let
$U=M_{m\times n}(K)$, $V=M_{n\times p}(K)$ and $W=M_{m\times p}(K)$
be the matrix spaces. The basis of $M_{a\times b}(K)$ is
$\{e_{ij}\mid 1\leq i\leq a,\, 1\leq j\leq b\}$, where $e_{ij}$ are the usual
matrix units. By $\{e'_{ij}\}$ we denote the dual basis of
$(M_{a\times b}(K))^\ast$, that is $(e'_{ij},e_{kl})=\delta_{ik}\delta_{jl}$.
Let $\mu:U\times V\lra W $ be the usual multiplication of matrices, then its
structure tensor equals
$$ \ov\mu=\sum_{i,j,k}e'_{ij}\otimes e'_{jk}\otimes e_{ik}, $$
where the sum is over all $1\leq i\leq m$, $1\leq j\leq n$, $1\leq k\leq p$.

Suppose that $\ov\mu$ can be represented as a sum of $r$ decomposable tensors:
$$ \ov\mu=\sum_{i=1}^r u_i\otimes v_i\otimes w_i, $$
$u_i\in U^\ast$ ($=M_{m\times n}(K)^\ast$), $v_i\in V^\ast$,
$w_i\in W$. Then there exists an algorithm, that computes the
product of $m\times n$ and $n\times p$ matrices over $R$ that
needs $r$ multiplications in $R$. For $a,b\in{\bf N}$ define {\em
convolution}
$$ (,)\::\: M_{a\times b}(R)\times M_{a\times b}(K)^\ast\lra R $$
by
$$(\sum r_{ij}e_{ij},\sum x_{ij}e'_{ij})=\sum x_{ij}r_{ij}, $$
where $r_{ij}\in R$, $x_{ij}\in K$, and the sums are taken over
$i=1,\ldots,a$, $j=1,\ldots, b$. Now it is possible to prove that for
arbitrary $A\in M_{m\times n}(R)$, $B\in M_{n\times p}(R)$
\begin{equation}    \label{f:2}
AB=\sum_{i=1}^r(A,u_i)(B,v_i)w_i\,.
\end{equation}

{\em Example}. Let $m=n=p=2$, then the tensor
$$ \ov\mu=\sum_{i,j,k=1,2} e'_{ij}\otimes e'_{jk}\otimes e_{ik} $$
can be represented as
\begin{eqnarray*}
\ov\mu &=& e'_{11}\otimes(e'_{12}+e'_{22})\otimes (e_{12}-e_{22})
+ (e'_{11}-e'_{12})\otimes e'_{22}\otimes (-e_{11}-e_{12}) \\
&+& (-e'_{21}+e'_{22})\otimes e'_{11}\otimes (-e_{21}-e_{22}) + e'_{22}\otimes(e'_{11}+e'_{21})\otimes(-e_{11}+e_{21}) \\
&+& (e'_{11}+e'_{22})\otimes(e'_{11}+e'_{22})\otimes (e_{11}+e_{22})
+ (e'_{11}+e'_{21})\otimes (-e'_{12}+e'_{11})\otimes (-e_{22}) \\
&+& (e'_{12}+e'_{22})\otimes (e'_{21}-e'_{22})\otimes e_{11}.
\end{eqnarray*}
Applying the described construction to this decomposition, we
obtain formula (\ref{f:1}).

The proof of the formula (\ref{f:2}) in the general case is left to
the interested reader.

Note that the tensor $\ov\mu$ is equivalent (under some Segre isomorphism)
to a certain tensor that can be represented in a rather symmetric form.

Denote $M_{ab}=M_{a\times b}(K)$, for brevity. Consider isomorphisms
$M_{mn}^\ast\lra M_{mn}$,
$M_{np}^\ast\lra M_{np}$, $M_{mp}\lra M_{pm}$ defined by $e'_{ij}\mapsto e_{ij}$,
$e'_{jk}\mapsto e_{jk}$, $e_{ik}\mapsto e_{ki}$ respectively. Let
$$ \fe: M^\ast_{mn}\otimes M^\ast_{np}\otimes M_{mp} \lra M_{mn}\otimes
M_{np}\otimes M_{pm} $$
be the corresponding Segre isomorphism. Then $\fe(\ov\mu)=S(m,n,p)$, where
$$ S(m,n,p)=\sum e_{ij}\otimes e_{jk}\otimes e_{ki}\,,$$
the sum is again over $1\leq i\leq m$, $1\leq j\leq n$, $1\leq k\leq p$.

Thus we have the following
\begin{prop}    \label{pr:rk}
The minimal number of multiplications needed to multiply $m\times n$
matrix (with non-commuting entries) by a $n\times p$ matrix equals
the rank of $S(m,n,p)$.
\end{prop}

This proposition is more or less known (I have no precise reference, but
cf. the exercises in the last section of the textbook \cite{Kos-Man}).

By application of the isomorphism $\fe$ to the decomposition for $\ov\mu$,
described in the example, we obtain decomposition
\begin{eqnarray}
S(2,2,2) &=& \sum_{i,j,k=1,2}e_{ij}\otimes e_{jk}\otimes e_{ki}=
e_{11}\otimes(e_{12}+e_{22})\otimes (e_{21}-e_{22}) \nonumber\\
&+& (e_{11}-e_{12})\otimes e_{22}\otimes (-e_{11}-e_{21}) + (-e_{21}+e_{22})
\otimes e_{11}\otimes (-e_{12}-e_{22}) \label{f:razls222} \\
&+& e_{22}\otimes(e_{11}+e_{21})\otimes(-e_{11}+e_{12}) + (e_{11}+e_{22})
\otimes(e_{11}+e_{22})\otimes (e_{11}+e_{22}) \nonumber\\
&+& (e_{11}+e_{21})\otimes (-e_{12}+e_{11})\otimes (-e_{22}) + (e_{12}+
e_{22})\otimes (e_{21}-e_{22})\otimes e_{11}. \nonumber
\end{eqnarray}

\paragraph{5. Group actions on algorithms.} Let $t\in V_1\otimes\ldots\otimes V_l$
be an arbitrary tensor. An {\em algorithm computing} $t$ is a set
$\{t_1,\ldots,t_n\}$ of decomposable tensors such that $t_1+\ldots+t_n=t$. For example,
\begin{eqnarray*}
{\cal S} &=& \{ e_{11}\otimes(e_{12}+e_{22})\otimes (e_{21}-e_{22}), \\
&& (e_{11}-e_{12})\otimes e_{22}\otimes (-e_{11}-e_{21}),\
(-e_{21}+e_{22})\otimes e_{11}\otimes (-e_{12}-e_{22}), \\
&& e_{22}\otimes(e_{11}+e_{21})\otimes(-e_{11}+e_{12}),\
(e_{11}+e_{22})\otimes(e_{11}+e_{22})\otimes (e_{11}+e_{22}), \\
&& (e_{11}+e_{21})\otimes (-e_{12}+e_{11})\otimes (-e_{22}),\
(e_{12}+e_{22})\otimes (e_{21}-e_{22})
\otimes e_{11}\ \}
\end{eqnarray*}
is an algorithm (which is, of course, called {\em Strassen algorithm}) computing $S(2,2,2)$.

It is evident that the extended isotropy group $\wt\Gamma(t)$ acts on the
set of all algorithms computing $t$ (in particular, on the set of all
optimal algorithms). The {\em automorphism group} of algorithm
$\{t_1,\ldots,t_n\}$ is the subgroup of all elements $\fe\in\wt\Gamma(t)$
preserving $\{t_1,\ldots,t_n\}$.

\begin{theorem} \label{th:Groote}
Let $t=S(2,2,2)$. Then the isotropy group $\Gamma(t)$ acts transitively on
the set of all optimal algorithms computing $t$.
\end{theorem}

This theorem is due to de Groote \cite{Groote}. It should be mentioned that
there are vague places in \cite{Groote}, in particular in the proofs of
Propositions 2.9 and 2.10; but the author found an independent proof of the theorem.

One of the main results of the present work is the following
\begin{theorem} \label{th:main}
The automorphism group of the Strassen algorithm is a finite group
isomorphic to $S_3\times S_3$.
\end{theorem}

(In fact, the concept of the ``automorphism group of an algorithm'' admits
some extension, which will be explained below).

\paragraph{6. Some authomophisms of the Strassen algorithm.} In this paragraph
we describe some authomorphisms of the Strassen algorithm. First we give a
formal proof, and then explain the origin of the athomorphisms.

Any authomorphism of the Strassen algorithm is, by definition, a Segre
authomorphism of the product $M_2\otimes M_2\otimes M_2$, where
$M_2=M_{22}=M_2(K)$.

\begin{prop}    \label{prop:AutS}
Let $T_1=\begin{pmatrix} 0 & -1 \\ 1 & -1 \end{pmatrix}$,
$T_2=\begin{pmatrix} 0 & 1 \\ 1 & 0 \end{pmatrix}$, and let $\Phi_i$
be the transformation of $M_2$ defined by $\Phi_i(x)=T_ixT_i^{-1}$. Define
$\rho:M_2\lra M_2$ by
$$\rho(\begin{pmatrix} a & b \\ c & d \end{pmatrix})
=\begin{pmatrix} d & -b \\ -c & a \end{pmatrix}.$$
Next, let $A_1$, $A_2$, $B_1$ and $B_2$ be the Segre authomorphisms
of the product $M_2\otimes M_2\otimes M_2$ defined by
$$ A_1(x\otimes y\otimes z)= y\otimes z\otimes x\,, $$
$$ A_2(x\otimes y\otimes z)= \rho(x)\otimes \rho(z)\otimes \rho(y)\,, $$
$$ B_i(x\otimes y\otimes z)= \Phi_i(x)\otimes \Phi_i(y)\otimes \Phi_i(z),
\quad i=1,2.$$

Then

(1) $A_1,A_2,B_1,B_2\in\Aut(\cS)$;

(2) $\lu A_1,A_2,B_1,B_2\ru = \lu A_1,A_2\ru \times \lu B_1,B_2\ru$
and $\lu A_1,A_2\ru\cong\lu B_1,B_2\ru\cong S_3$.
\end{prop}

{\em Proof.} First prove (2). Clearly, $A_1^3=1$ ($=\id_L$, where
we denote $L=M_2^{\otimes 3}$). Notice that $\rho^2=\id_{M_2}$,
whence $A_2^2=1$. Further,
\begin{eqnarray*}
(A_2A_1A_2)(x\otimes y\otimes z) &=& (A_2A_1)(\rho(x)\otimes
\rho(z)\otimes\rho(y))= A_2(\rho(z)\otimes\rho(y)\otimes\rho(x)) \\
&=& z\otimes x\otimes y=A_1^{-1}(x\otimes y\otimes z),
\end{eqnarray*}
whence $A_2A_1A_2=A_1^{-1}$. Hence $\lu A_1,A_2\ru\cong S_3$.

Next, $T_1$ and $T_2$ satisfy relations $T_1^3=T_2^2=1$, $T_2T_1T_2
=T_1^{-1}$; it follows that $\Phi_1$ and $\Phi_2$, as well as $B_1$
and $B_2$, satisfy the same relations, and therefore generate a group
isomorphic to $S_3$.

Since $\lu A_1,A_2\ru\cong S_3$, each nontrivial element of $\lu A_1,A_2\ru$
induces nontrivial permutation of the factors of the product
$M_2\otimes M_2\otimes M_2$. On the other hand, $\lu B_1,B_2\ru$
preserves each factor of the latter product. So $\lu A_1,A_2\ru
\cap\lu B_1,B_2\ru=1$.

It remains to prove that $A_i$ and $B_j$ commute for all $i$, $j$.
The fact that $A_1$ commutes with both $B_1$ and $B_2$ is almost
evident.

Next, it is easy to prove that $\rho$ commutes with both $\Phi_1$
and $\Phi_2$. (Check this, for example, for $\Phi_1$. Let
$x=\begin{pmatrix} a & b \\ c & d \end{pmatrix}$ be an arbitrary
matrix. Then
\\ $\Phi_1(x)=\begin{pmatrix} 0 & -1 \\ 1 & -1 \end{pmatrix} x
\begin{pmatrix} 0 & -1 \\ 1 & -1 \end{pmatrix}^{-1} =
\begin{pmatrix} 0 & -1 \\ 1 & -1 \end{pmatrix} \begin{pmatrix} a & b \\
c & d \end{pmatrix}\begin{pmatrix} -1 & 1 \\ -1 & 0 \end{pmatrix}=
\begin{pmatrix} c+d & -c \\ -a-b+c+d & a-c \end{pmatrix}$, \\
whence
$$ (\rho\Phi_1)(x)= \begin{pmatrix} a-c & c \\ a+b-c-d & c+d \end{pmatrix}; $$
on the other hand, $\rho(x)=\begin{pmatrix} d & -b \\ -c & a \end{pmatrix}$,
whence
$$ (\Phi_1\rho)(x)= \begin{pmatrix} a-c & c \\ -d+b-c+a & d+c \end{pmatrix}, $$
the same matrix.
The similar checking for $\Phi_2$ is even simpler.)

Now
$$(A_2B_i)(x\otimes y\otimes z)=A_2(\Phi_i(x)\otimes\Phi_i(y)
\otimes\Phi_i(z))=\rho(\Phi_i(x))\otimes\rho(\Phi_i(z))
\otimes\rho(\Phi_i(y));$$
on the other hand,
$$(B_iA_2)(x\otimes y\otimes z)=B_i(\rho(x)\otimes\rho(z)
\otimes\rho(y))=\Phi_i(\rho(x))\otimes\Phi_i(\rho(z))\otimes\Phi_i(\rho(y)),$$
which is the same, because of $\rho\Phi_i=\Phi_i\rho$.

Thus, (2) is proved.

Now we prove (1).

First note that $\cS$ is invariant under $A_1$. More precisely,  $\cS$
falls into three orbits
$$ \cO_0=\{(e_{11}+e_{22})\otimes (e_{11}+e_{22})\otimes(e_{11}+e_{22}) \}
= \{\delta\otimes\delta\otimes\delta\},$$
where $\delta=e_{11}+e_{22}=\begin{pmatrix} 1 & 0\\ 0 & 1\end{pmatrix}$,
\begin{eqnarray*}
\cO_1=\{ && e_{11}\otimes (e_{12}+e_{22})\otimes (e_{21}-e_{22}),\
\
(e_{21}-e_{22}) \otimes e_{11}\otimes (e_{12}+e_{22}), \\
&& (e_{12}+e_{22})\otimes (e_{21}-e_{22})\otimes e_{11}\ \},
\end{eqnarray*}
\begin{eqnarray*}
\cO_2=\{ && e_{22}\otimes (e_{11}+e_{21})\otimes
(-e_{11}+e_{12}),\ \
(-e_{11}+e_{12}) \otimes e_{22}\otimes (e_{11}+e_{21}), \\
&& (e_{11}+e_{21})\otimes (-e_{11}+e_{12})\otimes e_{22}\ \ \}.
\end{eqnarray*}
We call $\delta\otimes\delta\otimes\delta$ the {\em exceptional}
element of $\cS$, the other 6 elements are called {\em regular} ones.
The set of regular elements of $\cS$ will be denoted by $\cS_0$.

Further, $\rho$ takes matrices
$$ e_{11}=\begin{pmatrix} 1 & 0 \\ 0 & 0 \end{pmatrix}, \quad
e_{12}+e_{22}=\begin{pmatrix} 0 & 1 \\ 0 & 1 \end{pmatrix}, \quad
e_{21}-e_{22}=\begin{pmatrix} 0 & 0 \\ 1 & -1 \end{pmatrix} $$
to
$$ \begin{pmatrix} 0 & 0 \\ 0 & 1 \end{pmatrix}=e_{22}, \quad
\begin{pmatrix} 1 & -1 \\ 0 & 0 \end{pmatrix}=e_{11}-e_{12}, \quad
\begin{pmatrix} -1 & 0 \\ -1 & 0 \end{pmatrix}=-e_{11}-e_{21} $$
respectively, so $A_2$ takes tensor
$$e_{11}\otimes (e_{12}+e_{22})\otimes (e_{21}-e_{22})\in\cO_1$$
to
\begin{eqnarray*}
\rho(e_{11})\otimes \rho(e_{21}-e_{22})\otimes\rho(e_{12}+e_{22})
& = & e_{22}\otimes (-e_{11}-e_{21})\otimes (e_{11}-e_{12}) \\
&=& e_{22}\otimes (e_{11}+e_{21})\otimes (-e_{11}+e_{12})\in\cO_2\,.
\end{eqnarray*}

Since $A_2^2=1$ and $A_2$ normalizes $\lu A_1\ru$, it follows that
$A_2$ interchanges the orbits $\cO_1$ and $\cO_2$. Therefore,
$\lu A_1,A_2\ru $ preserves $\cO_1\cup\cO_2$ (and, moreover, acts
transitively on $\cO_1\cup\cO_2$). It is also evident that $A_2$
preserves $\delta\otimes\delta\otimes\delta$. Thus, $\cS$ is invariant
under $\lu A_1,A_2\ru$.

It is obvious that $\delta$ is invariant under $\Phi_1$ and $\Phi_2$, whence
$\delta\otimes\delta\otimes\delta$ is invariant under $B_1$ and $B_2$.
Next, it is easy to calculate that $\Phi_1$ takes matrices $e_{11}$,
$e_{12}+e_{22}$ and $e_{21}-e_{22}$ to $-e_{21}+e_{22}$, $e_{11}$ and
$-e_{12}-e_{22}$, respectively, and $\Phi_2$ takes them to $e_{22}$,
$e_{11}+e_{21}$ and $-e_{11}+e_{12}$, respectively. So $B_1$ takes
tensor $e_{11}\otimes (e_{12}+e_{22})\otimes (e_{21}-e_{22})$ to
$$(-e_{21}+e_{22})\otimes e_{11}\otimes(-e_{12}-e_{22})=(e_{21}-e_{22})
\otimes e_{11}\otimes(e_{12}+e_{22}),$$
and $B_2$ takes the same tensor
to $e_{22}\otimes (e_{11}+e_{21})\otimes (-e_{11}+e_{12})$.

Both of these tensors are in $\cO_1\cup\cO_2$. Taking into account
that $B_1$ and $B_2$ commute with $\lu A_1,A_2\ru$ and that the action
of $\lu A_1,A_2\ru$ on $\cO_1\cup\cO_2$ is transitive, we see that
both $B_1$ and $B_2$ preserve $\cO_1\cup\cO_2$. \hfill $\square$
\bigskip

Now we explain the origin of these automorphisms. First of all, one
sees immediately that $\cS$ is invariant under the cyclic shift (i.e.,
under $A_1$).

To construct the other automorphisms it is convenient to decompose
the product $M_2\otimes M_2\otimes M_2$ further. Let $V=\lu e_1,e_2\ru$
be the two-dimensional space and $V^\ast=\lu e^1,e^2\ru$ be its dual
space, $(e_i,e^j)=\delta_{ij}$. We identify $M_2$ with $V\otimes V^\ast$
by $e_{ij}\leftrightarrow e_i\otimes e^j$; then $M_2\otimes M_2\otimes M_2$
is identified with $V\otimes V^\ast\otimes V\otimes V^\ast\otimes V\otimes
V^\ast$. Then $\cS$ may be written as
$$ \{ e_1\otimes e^1\otimes(e_1+e_2)\otimes e^2\otimes e_2\otimes(e^1-e^2),\
e_1\otimes(e^1-e^2)\otimes e_2\otimes e^2\otimes(e_1+e_2)\otimes(-e^1), $$
$$ e_2\otimes(e^1-e^2)\otimes e_1\otimes e^1\otimes(e_1+e_2)\otimes e^2,\
e_2\otimes e^2\otimes(e_1+e_2)\otimes e^1\otimes e_1\otimes (-e^1+e^2)\,, $$
$$(e_1\otimes e^1+e_2\otimes e^2)\otimes (e_1\otimes e^1+e_2\otimes e^2) \otimes
(e_1\otimes e^1+e_2\otimes e^2), \
(e_1+e_2)\otimes e^1\otimes e_1\otimes(-e^1+e^2)\otimes e_2\otimes e^2,  $$
$$ (e_1+e_2)\otimes e^2\otimes e_2\otimes(e^1-e^2)\otimes e_1\otimes e^1\ \}. $$
The exceptional element equals $\delta\otimes\delta\otimes\delta$,
where $\delta=e_1\otimes e^1+e_2\otimes e^2$ is the ``identity tensor''
of the space $V\otimes V^\ast$. Moreover, each regular element of $\cS$
turns out to be decomposable, as an element of $(V\otimes V^\ast)^{\otimes 3}$.

Observe that for each regular element of $\cS$ its $V$-factors
are $e_1$, $e_2$ and $e_1+e_2$ (taken in various order and up to sign),
and the $V^\ast$-factors are $e^1$, $e^2$ and $e^1-e^2$.

Moreover, it is easy to check that the regular elements of $\cS$
are precisely the tensors of the form
$$ x_1\otimes x_2\otimes x_3\otimes x_4\otimes x_5\otimes x_6\,,$$
which satisfy the conditions

(1) $\{\lu x_1\ru,\lu x_3\ru,\lu x_5\ru\}=\{\lu e_1\ru,\lu e_2\ru,
\lu e_1+e_2\ru\}$, and $\{\lu x_2\ru,\lu x_4\ru,\lu x_6\ru\}=
\{\lu e^1\ru,\lu e^2\ru,\lu e^1-e^2\ru\}$;

(2) $(x_1,x_4)=(x_3,x_6)=(x_5,x_2)=0$;

(3) $(x_1,x_2)(x_3,x_4)(x_5,x_6)=-1$.

Observe that the triples of one-dimensional subspaces $\{\lu e_1\ru,
\lu e_2\ru,\lu e_1+e_2\ru\}$ and $\{\lu e^1\ru,
\lu e^2\ru,\lu e^1-e^2\ru\}$ are dual in the following sense: the
annihilator of an element of one of these triples is an element of
the other triple.

Now let $T$ be a linear transformation on $V$, preserving $\{\lu e_1\ru,
\lu e_2\ru,\lu e_1+e_2\ru\}$. Then the dual transformation $T^\ast$
of $V^\ast$ preserves $\{\lu e^1\ru, \lu e^2\ru,\lu e^1-e^2\ru\}$.
Moreover, it is evident that the transformation $\wt T=(T\otimes
T^\ast)^{\otimes 3}$ preserves the set of tensors satisfying
conditions (1), (2) and (3), i.e. it preserves $\cS_0$.
(It is clear also that $\wt T$ preserves the exceptional element
of $\cS$, because $T\otimes T^\ast$ preserves $\delta$).

It is clear that if $T'=\lambda T$, where $\lambda\in K^\ast$,
then $\wt{T'}=\wt T$. So $\wt T$ depends only on the permutation
induced by $T$ on $\{\lu e_1\ru,\lu e_2\ru,\lu e_1+e_2\ru\}$.
Taking as such a permutation a 3-cycle and a transposition on the
latter set, we obtain transformations $B_1=\wt T_1$ and $B_2=\wt T_2$,
described in the proposition \ref{prop:AutS}.

Finally, to construct automorphism $A_2$ we first construct a Segre
automorphism of the space $(V\otimes V^\ast)^{\otimes 3}$ that
interchanges $V$- and $V^\ast$-factors, and such that the corresponding
isomorphisms between $V$ and $V^\ast$ interchange triples
$\{\lu e_1\ru,\lu e_2\ru,\lu e_1+e_2\ru\}$ and $\{\lu e^1\ru,\lu e^2\ru,
\lu e^1-e^2\ru\}$. Namely, we define
$$ \fe: V\lra V^\ast, \quad \psi:V^\ast\lra V $$
by
$$ \fe:\quad e_1\mapsto e^2, \quad e_2\mapsto -e^1\,,\quad {\rm (whence}\
e_1+e_2\mapsto -e^1+e^2{\rm )}; $$
$$ \psi:\quad e^1\mapsto e_2, \quad e^2\mapsto -e_1\,,\quad
{\rm (whence}\  e^1-e^2\mapsto e_1+e_2\,{\rm )}. $$ (thus, $\fe$
and $\psi$ maps each of the one-dimensional subspaces $\lu
e_1\ru$, $\lu e_2\ru$, $\lu e_1+e_2\ru$, $\lu e^1\ru$, $\lu
e^2\ru$, and $\lu e^1-e^2\ru$ to its annihilator). Next, we put
$\rho(x\otimes y)=\psi(y)\otimes\fe(x)$. Then it is easy to see
that $\rho(\delta)=\delta$. Now one can guess that the map defined
by
$$ x\otimes y\otimes z\mapsto \rho(x)\otimes\rho(z)\otimes\rho(y) $$
leaves the set of tensors satisfying (1),(2) and (3) invariant.

\paragraph{7. Additional symmetries.} Observe that any element of
$\Aut(\cS)$ is a Segre authomorphism not only with respect to the decomposition
$M_2\otimes M_2\otimes M_2$, but with respect to the decomposition
$V\otimes V^\ast\otimes V\otimes V^\ast\otimes V\otimes V^\ast$
also. There are, however, symmetries of the Strassen algorithm
that are Segre authomorphisms with respect to $V\otimes V^\ast\otimes \ldots
\otimes V^\ast$, but not with respect to $M_2\otimes M_2\otimes M_2$.

The origin of these symmetries, informally speaking, is the following.
Denote by $\cS_0$ the set of all regular elements of $\cS$. Write the
decomposition
\begin{equation}    \label{f:razl_S0}
S(2,2,2)=\delta\otimes\delta\otimes\delta +\sum\cS_0\,,
\end{equation}
where $\sum \cS_0$ is the sum of all elements of $\cS_0$.

But the tensors $S(2,2,2)$ and $\delta\otimes\delta\otimes\delta$
are actually of the same kind, in the following sense. Let $(i_1,i_2,i_3)$
be any permutation of $(2,4,6)$. Then $\{\{1,i_1\}, \{3,i_2\},\{5,i_3\}\}$
is a partition of the set $\{1,2,3,4,5,6\}$ into pairs such
that every pair contains an even number and an odd one (and,
conversely, any such partition can be obtained in the described way).
The partititon $\{\{1,i_1\}, \{3,i_2\},\{5,i_3\}\}$ determines a
partition in pairs of the form $\{V,V^\ast\}$ of the factors of the product
$V\otimes V^\ast\otimes \ldots \otimes V^\ast$. Next, we take in
each of the three factors $V\otimes V^\ast$ the identity tensor
$\delta$, and form the product of these three tensors. The tensor
obtained in this way we denote by $t_{i_1i_2i_3}$. For example,
\begin{eqnarray*}
t_{462} &=& e_1\otimes e^1\otimes e_1\otimes e^1\otimes e_1\otimes e^1 +
e_1\otimes e^2\otimes e_1\otimes e^1\otimes e_2\otimes e^1 \\
&+& e_1\otimes e^1\otimes e_2\otimes e^1\otimes e_1\otimes e^2 +
e_1\otimes e^2\otimes e_2\otimes e^1\otimes e_2\otimes e^2 \\
&+& e_2\otimes e^1\otimes e_1\otimes e^2\otimes e_1\otimes e^1 +
e_2\otimes e^2\otimes e_1\otimes e^2\otimes e_2\otimes e^1 \\
&+& e_2\otimes e^1\otimes e_2\otimes e^2\otimes e_1\otimes e^2 +
e_2\otimes e^2\otimes e_2\otimes e^2\otimes e_2\otimes e^2\,.
\end{eqnarray*}
Then, clearly, $\delta\otimes\delta\otimes\delta$ is nothing else
but $t_{246}$. Moreover,
$$ S(2,2,2)=\sum_{i,j,k=1,2}e_{ij}\otimes
e_{jk}\otimes e_{ki}=\sum_{i,j,k=1,2} e_i\otimes e^j\otimes e_j\otimes
e^k\otimes e_k\otimes e^i=t_{624}. $$
Thus, we see that the tensors $S(2,2,2)$ and $\delta\otimes\delta\otimes\delta$
are of the same kind indeed and differ only in permutation of factors.

Now we give a formal definition. The {\em extended Strassen algorithm}
is the set
$$\wt\cS=\{-S(2,2,2), \delta\otimes\delta\otimes\delta\}\cup\cS_0\,.$$
(Note that the sum of the elements of $\wt\cS$ is the zero tensor).
An {\em authomorphism} of $\wt\cS$ is a Segre authomorphism of $V\otimes
V^\ast\otimes \ldots\otimes V^\ast$ that leaves $\wt\cS$ invariant.

\begin{prop}    \label{pr:AutwtS}
Let $T_i\in GL(V)$, $T^\ast_i\in GL(V^\ast)$, $\fe:V\lra V^\ast$ and
$\psi:V^\ast\lra V$ be defined by
$$ T_1\::\:e_1\mapsto e_2, \quad e_2\mapsto -e_1-e_2\,;\quad
T_1^\ast\::\:e^1\mapsto -e^1+e^2, \quad e^2\mapsto -e^1\,;$$
$$ T_2\::\:e_1\mapsto e_2, \quad e_2\mapsto e_1\,;\quad  T_2^\ast
\::\:e^1\mapsto e^2, \quad e^2\mapsto e^1\,;$$
$$\fe\::\: e_1\mapsto e^2,\quad e_2\mapsto -e^1;\quad \psi\::\:
e^1\mapsto e_2,\quad e^2\mapsto -e_1\,.$$
Let $\wt A_2$, $\wt A_3$, $\wt B_1$, $\wt B_2$ be Segre authomorphisms
of the product $V\otimes V^\ast\otimes V\otimes V^\ast\otimes V\otimes V^\ast$
defined by
$$ \wt A_2(x_1\otimes y_1\otimes x_2\otimes y_2\otimes x_3\otimes y_3)=
\psi(y_1)\otimes\fe(x_1)\otimes \psi(y_3)\otimes\fe(x_3)\otimes \psi(y_2)
\otimes\fe(x_2), $$
$$ \wt A_3(x_1\otimes y_1\otimes x_2\otimes y_2\otimes x_3\otimes y_3)=
-x_1\otimes y_3\otimes x_3\otimes y_2\otimes x_2\otimes y_1\,,$$
$$\wt B_i(x_1\otimes y_1\otimes x_2\otimes y_2\otimes x_3\otimes y_3)=
T_i(x_1)\otimes T_i^\ast(y_1)\otimes T_i(x_2)\otimes T_i^\ast(y_2)\otimes
T_i(x_3)\otimes T_i^\ast(y_3), \quad i=1,2. $$
Then

(1) $\lu\wt A_2,\wt A_3,\wt B_1,\wt B_2\ru =\lu\wt A_2,\wt A_3\ru
\times\lu\wt B_1,\wt B_2\ru$, $\lu\wt A_2,\wt A_3\ru\cong D_{12}$
(the dihedral group of order $12$), and $\lu\wt B_1,\wt B_2\ru\cong S_3$;

(2) $\wt A_2,\wt A_3,\wt B_1,\wt B_2\in\Aut(\wt\cS)$.
\end{prop}

{\em Proof.} We use results of Proposition~\ref{prop:AutS}. Identify
$V\otimes V^\ast\otimes \ldots\otimes V^\ast$ with $M_2\otimes M_2\otimes M_2$.
Then $\wt B_1$, $\wt B_2$ and $\wt A_2$ correspond to $B_1$, $B_2$ and $A_2$ of
Proposition~\ref{prop:AutS}. So $\lu\wt B_1,\wt B_2\ru\cong S_3$,
$\wt A_2^2=1_L$ and $\wt B_1$, $\wt B_2$ commute with $\wt A_2$.
Next, it is almost evident that $\wt A_3$ commutes with $\wt B_1$ and
$\wt B_2$. So the subgroups $\lu\wt A_2,\wt A_3\ru$
and $\lu\wt B_1,\wt B_2\ru$ commute elementwise.

Note that $\fe$ and $\psi$ are inverse to each other, up to $-1$:
$\psi\fe=-1_V$, $\fe\psi=-1_{V^\ast}$.

It is clear that $\wt A_3^2=1_L$.

Consider the product $\wt A_2\wt A_3$. We have
\begin{eqnarray*}
(\wt A_2\wt A_3)(x_1\otimes y_1\otimes x_2\otimes y_2\otimes
x_3\otimes y_3) &=& \wt A_2(-x_1\otimes y_3\otimes x_3\otimes y_2\otimes
x_2\otimes y_1) \\
&=& -\psi(y_3)\otimes\fe(x_1)\otimes \psi(y_1)\otimes\fe(x_2)
\otimes\psi(y_2)\otimes\fe(x_3),
\end{eqnarray*}
whence
\begin{eqnarray*}
&& (\wt A_2\wt A_3)^2(x_1\otimes y_1\otimes x_2\otimes y_2\otimes
x_3\otimes y_3)\\
&&=\psi\fe(x_3)\otimes\fe\psi(y_3)\otimes \psi\fe(x_1)\otimes
\fe\psi(y_1)\otimes\psi\fe(x_2) \otimes\fe\psi(y_2)\\
&&= (-x_3)\otimes(-y_3)\otimes (-x_1)\otimes(-y_1)\otimes(-x_2)\otimes(-y_2) \\
&& =x_3\otimes y_3\otimes x_1\otimes y_1\otimes x_2\otimes y_2.
\end{eqnarray*}
Set $\wt A_1=(\wt A_2\wt A_3)^2$. Then $\wt A_1^3=1$, whence $(\wt A_2\wt A_3)^6=1$.
It is also obvious that $\wt A_1$ corresponds to $A_1^{-1}$ under the
identification described above.

The transformations $\wt A_2$ and $\wt A_3$ satisfy the defining
relations of the group $D_{12}$, so $\lu\wt A_2,\wt A_3\ru$ is a
homomorphic image of $D_{12}$. On the other hand,
the group of permutations of the factors of the product $V\otimes
V^\ast\otimes \ldots \otimes V^\ast$, induced by $\lu\wt A_2,\wt A_3\ru$,
is isomorphic to $D_{12}$. So $\lu\wt A_2,\wt A_3\ru\cong D_{12}$, and
moreover $1$ is the only element of $\lu\wt A_2,\wt A_3\ru$
that induces the trivial permutation of the factors. Since $\lu\wt B_1,\wt B_2\ru$
preserves each factor, it follows that $\lu\wt A_2,\wt A_3\ru\cap
\lu\wt B_1,\wt B_2\ru=1$, whence
$$\lu\wt A_2,\wt A_3,\wt B_1,\wt B_2\ru =\lu\wt A_2,\wt A_3,
\ru\times\lu\wt B_1,\wt B_2\ru.$$

Thus, (1) is proved. Prove (2).

The transformations $\wt B_1$, $\wt B_2$ and $\wt A_2$ preserve $\cS$ by
Proposition~\ref{prop:AutS} and the identification. So they preserve $\wt\cS$ also.
Next, observe that $\wt A_3$ takes element
$$e_1\otimes e^1\otimes (e_1+e_2)\otimes e^2\otimes e_2\otimes(e^1-e^2)\in\cS_0$$
to
$$-e_1\otimes(e^1-e^2)\otimes e_2\otimes e^2\otimes(e_1+e_2)\otimes e^1\in \cS_0\,.$$
Taking into account that $\wt A_3$ commutes with $\lu\wt B_1,\wt B_2\ru$ and that
$\lu\wt B_1,\wt B_2\ru$ is transitive on $\cS_0$, we see that $\wt A_3$ leaves $\cS_0$
invariant.

Finally, it is easy to see that $\wt A_3$ interchanges $-S(2,2,2)$ and
$\delta\otimes\delta\otimes\delta$. Thus, $\wt A_3$ preserves $\wt\cS$.
\hfill $\square$

\paragraph{8. The full authomorphism groups of $\cS$ and $\wt\cS$.}
In this paragraph we show that the authomorphism groups for $\cS$
and $\wt\cS$, described in Propositions \ref{prop:AutS} and \ref{pr:AutwtS},
are actually
the full authomorphism groups $\Aut(\cS)$ and $\Aut(\wt\cS)$.

First observe the following. Let $U\otimes V$ be the tensor product of two
spaces, and let $L\sse U\otimes V$ be a subspace. Then there exists the
least subspace $X\sse U$ such that $L\sse X\otimes V$. We call this $X$
the {\em quasiprojection} of $L$, and we use notation $X=\qpr_U(L)$.
The quasiprojection can be easily found. Namely, let $x_1,\ldots,x_l\in L$
be a basis of $L$, $v_1,\ldots, v_n$ be a basis of $V$, and let
$u_{ij}\in U$, where $1\leq i\leq l$, $1\leq j\leq n$, be the elements
such that $x_i=\sum_{1\leq j\leq n} u_{ij}\otimes v_j$. Then
$$ \qpr_U(L)=\lu u_{ij}\mid 1\leq i\leq l,\ 1\leq j\leq n\ru. $$

If $t\in U\otimes V$ is a decomposable tensor, then $\qpr_U(\lu t\ru)$
is a one-dimensional subspace. Generally, if $L=\lu t\ru$ is a one-dimensional
subspace, then $\dim \qpr_U(\lu t\ru)=\rk(t)$. Moreover, for any two
subspaces $L_1,L_2\sse U\otimes V$ we have $\qpr_U(L_1+L_2)=
\qpr_U(L_1)+\qpr_U(L_2)$.

Below we need a simple lemma, whose proof is left to the reader.
\begin{lemma}   \label{l:scalar}
Let $V$ be a space over an infinite field $K$ and let $L_1,\ldots,L_l\sse V$
be one-dimensional subspaces. Suppose that $\lu L_1,\ldots,L_l\ru=V$
and that there is no partititon $\{L_1,\ldots,L_l\}=A\sqcup B$ such
that $\lu A\ru\cap\lu B\ru=0$. Let $\fe\in GL(V)$ be a linear transformation
such that $\fe(L_i)=L_i$ for all $i=1,\ldots,l$. Then $\fe=\lambda E$
is a multiplication by a scalar. \hfill $\square$
\end{lemma}
\begin{prop}    \label{pr:AutSfull}
Let $H=\lu A_1,A_2,B_1,B_2\ru$ be the subgroup of $\Aut(\cS)$ described in
Proposition~\ref{prop:AutS}. Then $H=\Aut(\cS)$.
\end{prop}

{\em Proof.} Let $L=M_2\otimes M_2\otimes M_2$; denote the factors by
$X_1$, $X_2$, $X_3$:
$$ L=M_2\otimes M_2\otimes M_2=X_1\otimes X_2\otimes X_3\,.$$

Let $N\sse\Aut(\cS)$ be the subgroup of all elements that preserve
each factor. We know from Proposition~\ref{prop:AutS} (and its proof) that
any permutation of factors is induced by a suitable element of
$\lu A_1\,,A_2\ru$, and that the latter subgroup acts on
$\{X_1\,,X_2\,,X_3\}$ exactly; on the other hand, $\lu B_1\,,B_2\ru$
preserves each factor. So it is sufficient (and necessary) to
prove that $N=\lu B_1\,,B_2\ru$.

We number the elements of $\cS$ as follows:
$$ s_1=e_{11}\otimes(e_{12}+e_{22})\otimes (e_{21}-e_{22}),\quad
s_2=(-e_{11}+e_{12})\otimes e_{22}\otimes (e_{11}+e_{21}),$$
$$s_3=(e_{21}-e_{22})\otimes e_{11}\otimes (e_{12}+e_{22}), \quad
s_4=e_{22}\otimes(e_{11}+e_{21})\otimes(-e_{11}+e_{12}),$$
$$ s_5=(e_{11}+e_{21})\otimes (-e_{11}+e_{12})\otimes e_{22}\,,\quad
s_6=(e_{12}+e_{22})\otimes (e_{21}-e_{22})
\otimes e_{11}\,,$$
$$ s_7=(e_{11}+e_{22})\otimes(e_{11}+e_{22})\otimes (e_{11}+e_{22}) \
(=\delta\otimes\delta\otimes\delta). $$

Evidently, $N$ induces a subgroup, which will be denoted by $N_i$,
of the projective group $PGL(X_i)$.

Let $C$ be the set of quasiprojections of elements of $\cS$ to
$X_1$:
$$ C=\{\qpr_{X_1}(\lu x\ru)\mid x\in\cS\}. $$
It is obvious that $N_1$ preserves $C$. We can write $C$ explicitly:
$C=\{r_1,r_2,\ldots,r_7\}$, where $r_1=\lu e_{11}\ru$, $r_2=\lu e_{11}-e_{12}\ru$,
$r_3=\lu e_{21}-e_{22}\ru$, $r_4=\lu e_{22}\ru$, $r_5=\lu e_{11}+e_{21}\ru$,
$r_6=\lu e_{12}+e_{22}\ru$, and $r_7=\lu e_{11}+e_{22}\ru$.

We say that a set of subspaces $L_1,\ldots,L_t$ of a vector space is
{\em independent}, if $\lu L_1,\ldots,L_t\ru=L_1\oplus\ldots\oplus L_t$;
otherwise it is {\em dependent}. The following three triples of subspaces
$r_i$ are dependent: $\{r_1,r_4,r_7\}$, $\{r_2,r_6,r_7\}$, $\{r_3,r_5,r_7\}$.
Any other triple $\{r_i,r_j,r_k\}$ is independent.

Clearly, $N_1$ permutes subspaces $r_i$ and preserves the dependency
relation. So $N_1$ fixes $r_7$ and preserves the set of three pairs
$\{r_1,r_4\}$, $\{r_2,r_6\}$, and $\{r_3,r_5\}$. It follows that $N$
fixes $s_7$ and preserves $\{\{s_1,s_4\},\{s_2,s_6\},\{s_3,s_5\}\}$.
Similarly one can prove, considering quasiprojections to $X_2$ and $X_3$,
that $N$ preserves $\{\{s_1,s_5\},\{s_2,s_3\},\{s_4,s_6\}\}$ and
$\{\{s_1,s_2\},\{s_3,s_4\},\{s_5,s_6\}\}$.

Let $\Phi_1$ and $\Phi_2$ be the transformations described in
Proposition~\ref{prop:AutS}, and $\ov\Phi_i$ be their images in the projective
group. Then $\ov\Phi_{1,2}$ permute pairs $\{r_1,r_4\}$, $\{r_2,r_6\}$,
and $\{r_3,r_5\}$ as follows:
$$ \ov\Phi_1:\ \{r_1,r_4\}\mapsto\{r_2,r_6\}\mapsto\{r_3,r_5\}
\mapsto\{r_1,r_4\}, $$
$$ \ov\Phi_2:\ \{r_1,r_4\}\mapsto\{r_1,r_4\},\quad \{r_2,r_6\}
\leftrightarrow\{r_3,r_5\}. $$
Therefore, any permutation of $\{\{r_1,r_4\},\{r_2,r_6\},\{r_3,r_5\}\}$
is induced by an element of $\lu\ov\Phi_1,\ov\Phi_2\ru$. So any permutation
of the set of three pairs $\{\{s_1,s_4\},\{s_2,s_6\},\{s_3,s_5\}\}$
is induced by an element of $\lu B_1,B_2\ru$.

Now we can prove the proposition. Let $g\in N$, and let $g_1\in\lu B_1,B_2\ru$
be an element such that the permutations of $\{\{s_1,s_4\},\{s_2,s_6\},\{s_3,s_5\}\}$,
induced by $g$ and $g_1$, coincide. Then the element $g_2=g_1^{-1}g$
is in $N$, preserves each of the three pairs $\{s_1,s_4\}$, $\{s_2,s_6\}$,
and $\{s_3,s_5\}$, and, moreover, leaves invariant two sets
$\{\{s_1,s_5\},\{s_2,s_3\},\{s_4,s_6\}\}$ and
$\{\{s_1,s_2\},\{s_3,s_4\},\{s_5,s_6\}\}$. It easily follows that
$g_2$ fixes each $s_i$. Now let $i=1,2$ or $3$, and let $h_i$ be the
image of $g_2$ in $PGL(X_i)$. Then $h_i$ preserves each of the
subspaces $r_1,\ldots,r_7$, whence $h_i=1$ by Lemma~\ref{l:scalar}. Therefore
$g_2$ is a scalar also. Since $g_2$ preserves $S(2,2,2)$, it follows
that $g_2=1$, so $g=g_1\in\lu B_1,B_2\ru$. \hfill $\square$

\begin{prop}    \label{pr:Aut_wtS_full}
Let $H=\lu\wt A_2,\wt A_3,\wt B_1,\wt B_2\ru$ be the subgroup of
$\Aut(\wt\cS)$ described in Proposition~\ref{pr:AutwtS}. Then $H=\Aut(\wt\cS)$.
\end{prop}

{\em Proof.} We give a sketch of a proof, leaving some details to the reader.

Number the factors of the product $V\otimes V^\ast\otimes\ldots\otimes V^\ast$
as
$$ V\otimes V^\ast\otimes V\otimes V^\ast\otimes V\otimes V^\ast=
X_1\otimes X_2\otimes X_3\otimes X_4\otimes X_5\otimes X_6\,.$$

We have $\wt\cS=\{u_1,u_2\}\cup\cS_0$, where $u_1=-S(2,2,2)$ and
$u_2=\delta^{\otimes3}$. Observe that for $t\in\cS_0$ the quasiprojection
$\qpr_{X_i}(\lu t\ru)$ is of dimension 1, for all $i=1,\ldots,6$. On the
other hand, if $t\in\{u_1,u_2\}$, then $\qpr_{X_i}(\lu t\ru)=X_i$. Hence
$\Aut(\wt\cS)$ leaves $\{u_1,u_2\}$ invariant.

Let $1\leq i\ne j\leq6$. It is easy to see that $\qpr_{X_i\otimes X_j}
(\lu\delta\otimes\delta\otimes\delta\ru)$ is one-dimensional if
$\{i,j\}=\{1,2\}$, $\{3,4\}$ or $\{5,6\}$, and is equal to the full space
$X_i\otimes X_j$ otherwise. Similarly the quasiprojection
$\qpr_{X_i\otimes X_j}(\lu S(2,2,2)\ru)$ is one-dimensional if
$\{i,j\}=\{2,3\}$, $\{4,5\}$ or $\{1,6\}$ and is equal to $X_i\otimes X_j$
otherwise. It follows that if $g\in\Aut(\wt\cS)$ and $\pi$ is a
permutation of $\{1,\ldots,6\}$ (or, more precisely, of $\{X_1,\ldots,
X_6\}$), induced by $g$, then $\pi$ preserves the set of six pairs
$\{\{1,2\}, \{2,3\},\{3,4\},\{4,5\},\{5,6\},\{6,1\}\}$. Therefore
$\pi\in D$, where $D\cong D_{12}$ is the group of permutations of
$\{1,\ldots,6\}$ that preserve or invert the cyclic order $(1,2,\ldots,6)$.

Note, however, that any element of $D$ is induced by an element of
$\lu\wt A_2,\wt A_3\ru$. Hence $\Aut(\wt\cS)=N\lu\wt A_2,\wt A_3\ru$,
where $N\leq\Aut(\wt\cS)$ is the subgroup of all elements preserving each
$X_i$. So it is sufficient to prove that $N=\lu\wt B_1,\wt B_2\ru$.

Let $g\in N$, $g=g_1\otimes\ldots\otimes g_6$, $g_i\in GL(X_i)$
(note that $g_i$ is determined up to multiplication by a scalar).
Obviously, $g_1$ preserves the set of quasiprojections
$$ C=\{\qpr_{X_1}(\lu t\ru)\mid t\in\cS_0\}.$$
This $C$ consists of three subspaces $\lu e_1\ru$, $\lu e_2\ru$,
$\lu e_1+e_2\ru$. For any permutation $\pi$ of $C$ there exists
an element $h\in\lu\wt B_1,\wt B_2\ru$, $h=h_1\otimes\ldots\otimes h_6$
such that $h_1$ acts on $C$ by $\pi$. Therefore, there exist
elements $x=x_1\otimes\ldots\otimes x_6\in N$ and $h\in\lu\wt B_1,\wt B_2\ru$
such that $g=xh$ and $x_1$ acts on $C$ trivially. So it is sufficient
to show that if $x=x_1\otimes\ldots\otimes x_6\in N$ and $x_1$ acts on $C$ trivially,
then $x=1$.

Since $x_1$ fixes each of the three spaces $\lu e_1\ru$, $\lu e_2\ru$,
and $\lu e_1+e_2\ru$, we see that $x_1$ is a scalar. Next, since $x$
preserves $\{u_1,u_2\}$, the quasiprojection of $\lu u_1\ru$ to
$X_1\otimes X_2$ is $X_1\otimes X_2$, and the quasiprojection of
$\lu u_2\ru$ to this factor is one-dimensional, it follows that
$xu_1=u_1$ and $xu_2=u_2$. In particular, $x(\delta\otimes\delta\otimes
\delta)=\delta\otimes\delta\otimes\delta$. Hence $(x_1\otimes x_2)\delta
=\lambda\delta$, where $\lambda\in K^\ast$. So $x_1$ and $x_2$ are
dual maps, up to multiplication by a scalar. Similarly $x_4=x_3^\ast$
and $x_6=x_5^\ast$ up to scalar. Furthermore, as $x$ preserves
$u_1=-S(2,2,2)$, we see that $x_3=x_2^\ast$, $x_5=x_4^\ast$, and
$x_6=x_1^\ast$, up to scalar. Therefore, $x_1=x_3=x_5$ (up to scalar)
and $x_2=x_4=x_6=x_1^\ast$. Since $x_1$ is a scalar, the other $x_i$
are scalars also, whence $x=\mu$ is a scalar. Finally, since
$xu_1=u_1$ and $u_1\ne0$, we get $\mu=1$, that is $x=1$.
\hfill $\square$

\paragraph{Acknowledgements.} I would like to thank A.E.Zalesskii,
A.S.Kleshchev and Pham Huu Tiep for their friendly support during
my work on this text.

\end{document}